\documentclass{article}%
\usepackage{amsmath}
\usepackage{graphicx}%
\usepackage{amsfonts}%
\usepackage{amssymb}
\begin{document}

\title{Ambiguities of the Seiberg-Witten map in the presence of matter field}
\author{Bing Suo\thanks{e-mail: bbs@phy.nwu.edu.cn},
Pei Wang\thanks{e-mail:pwang@phy.nwu.edu.cn} { } and
Liu Zhao\thanks{e-mail: lzhao@phy.nwu.edu.cn}
\\Institute of Modern Physics, Northwest University, Xian 710069, China}
\date{}
\maketitle

\begin{abstract}
The ambiguities of the Seiberg-Witten map for gauge field coupled with fermionic
matter are discussed. We find that only part of the ambiguities can be
absorbed by gauge transformation and/or field redefinition and thus are negligible.
The existence of matter field makes some other part of the ambiguities difficult
to be absorbed by gauge transformation or field redefinition.

\end{abstract}



\section{Introduction}

Recently, quantum field theories on noncommutative spaces have received
considerable interests.Though historically the concept of space(time)
noncommutativity was proposed over 50 years ago\cite{1}, most serious study
and progress in field theories defined on such spaces was made within the last
few years\cite{2,3,4}.The reason that there has been a renewed interests in
noncommutative spaces is due to the studies in string theory\cite{5,6}. It was
found that in the presence of a NS-NS background $B$ field, the quantization
of open string would result in spacial noncommutativity at the boundary, i.e.
on the $D$-branes. Among all the important achievements in the study of
noncommutative field theories, e.g. existence and construction of
nonperturbative solutions (solitons, instantons, monopoles etc.), relation to
$D$-brane and tachyon condensation and so on, the so-called Seiberg-Witten
map\cite{6} is a particularly useful idea for exploring various properties of
noncommutative field theories and thus an important subject of intensive study.

In their original paper\cite{6}, based on the observation that different
regularization schemes (point splitting vs. Pauli-Villars) in the field theory
limit of string theory leads to either a commutative or a noncommutative field
theory, Seiberg and Witten proposed a map which establish an equivalent
relationship between an ordinary gauge theory and a noncommutative one. They
then argued that the same kind of equivalent relationship should also hold
between noncommutative gauge theories with a small deviation in the
noncommutative parameter $\theta$. Many authors subsequently studied various
properties of Seiberg-Witten map\cite{7,8,9,10}, including applying
Seiberg-Witten map to different gauge theories, trying to find the accumulated
effect of successive Seiberg-Witten maps (i.e. solving the so-called
Seiberg-Witten equation)\cite{11,12}, and so on. Some paper revealed that fact
that, in general, Seiberg-Witten maps between noncommutative gauge theories of
different noncommutative parameters possess ambiguities. Ref.\cite{13}, argued
that part of the ambiguities of Seiberg-Witten map between noncommutative
gauge theories can be absorbed by a gauge transformation, while the rests of
the ambiguities can be removed by a field redefinition. In this paper, we
would like to re-exam the problem of ambiguities of Seiberg-Witten map.
However, in contrast to ref.\cite{13}, the field theories we consider are
noncommutative gauge theories coupled to some noncommutative matter fields
(represented by some spinor field in noncommutative space). As we shall see,
Seiberg-Witten maps between noncommutative gauge theories coupled to matter
fields would also yield some ambiguities, and, although some part of the
ambiguities could be absorbed by a gauge transformation, there are some other
parts which can neither be absorbed by gauge transformations nor by field
redefinitions. In other words, the presence of matter fields spoils the
consistency of the kind of argument made in \cite{13}, (though it seems that
without matter field the argument of ref. \cite{13} makes perfect sense), and
therefore the problem of cancellation of ambiguities in Seiberg-Witten map
should be considered as still an open question.

\section{Seiberg-Witten map for pure gauge theories: a brief review}

Before going into the details of Seiberg-Witten map for gauge theories coupled
with matter, we would like to first make a brief review on the works on
Seiberg-Witten maps for a purely gauge theory. This would not only provide the
theoretical setup for our subsequent work, but also fix the notations.

By definition, a noncommutative gauge theory is gauge theory with gauge
potential depending on noncommutative space coordinates (spacetime
noncommutativity is not considered in most cases). In practice, one often uses
the so-called Moyal product algebra of ordinary functions on commutative
spaces to represent the algebra of functions on noncommutative spaces. In this
formalism, the gauge field strength for a noncommutative gauge theory can be
written as%
\[
\hat{F}_{ij}=\partial_{i}\hat{A}_{j}-\partial_{i}\hat{A}_{i}-i[\hat{A}_{i}%
,\hat{A}_{j}]_{\ast},
\]
where throughout this paper we use hatted notations to represent
noncommutative field objects. In the above equation, the $\ast$-commutator is
defined as%
\[
\lbrack\hat{A},\hat{B}]_{\ast}=\hat{A}\ast\hat{B}-\hat{B}\ast\hat{A},
\]
where for any pair of ordinary functions $f(x)$ and $g(x)$, the Moyal product
$\ast$ is given as\cite{14}%
\[
f\ast g(x)=e^{\frac{i}{2}\theta^{ij}\frac{\partial}{\partial\xi^{i}}%
\frac{\partial}{\partial\varsigma^{j}}}f(x+\xi)g(x+\varsigma)|_{\xi
=\varsigma=0}=fg+\frac{i}{2}\theta^{ij}\partial_{i}f\partial_{j}g+O(\theta
^{2}).
\]
The parameter $\theta$ is hence forth referred to as noncommutativity
parameter. The noncommutative gauge transformations take the same form as
ordinary gauge transformations, the only difference is that one has to replace
ordinary products between functions by Moyal products:%
\begin{align}
\delta_{\hat{\lambda}}\hat{A}_{i}  &  =\partial_{i}\hat{\lambda}%
+i[\hat{\lambda},\hat{A}_{i}],\nonumber\\
\delta_{\hat{\lambda}}\hat{F}_{ij}  &  =i[\hat{\lambda},\hat{F}_{ij}],
\label{gaugefield}%
\end{align}
where $\hat{\lambda}$ is an infinitesimal noncommutative gauge parameter.

The ingenious observation made by Seiberg and Witten on noncommutative gauge
theories can be stated in the following simple sentence: \emph{there exists an
equivalent relation between noncommutative gauge theories of noncommutativity
parameters }$\theta$ \emph{and }$\theta+\delta\theta$\emph{, where the
equivalence is established by identifying the gauge equivalent classes of both
theories. }In terms of mathematical formalism, the last statement can be
written as%
\begin{equation}
\tilde{A}_{i}(\hat{A})+\delta_{\tilde{\lambda}}\tilde{A}_{i}%
(\hat{A})=\tilde{A}_{i}(\hat{A}+\delta_{\hat{\lambda}}\hat{A}), \label{SW}%
\end{equation}
or more compactly,%
\begin{equation}
\delta_{\tilde{\lambda}}\delta\hat{A}=\delta\delta_{\tilde{\lambda}}\hat{A},
\label{compact}%
\end{equation}
where $\tilde{A},\tilde{\lambda}$ are respectively the gauge potential and
gauge parameter in the noncommutative gauge theory with noncommutativity
parameter $\theta+\delta\theta$, and the operator $\delta$ in (\ref{compact})
represents the variation with respect to $\theta$. One thing one has to take
notice is that the variation $\delta$ not only acts on functions but also on
Moyal products:%
\begin{equation}
\delta(f\ast g)=\delta f\ast g+f\ast\delta g+\frac{i}{2}\delta\theta
^{ij}\partial_{i}f\ast\partial_{j}g. \label{var}%
\end{equation}

In order to solve (\ref{SW}) or (\ref{compact}), we write
$\tilde{A}=\hat{A}+\delta\hat{A}(\hat{A},\hat{\lambda},\delta\theta
)+O(\delta\theta^{2}),\tilde{\lambda}=\hat{\lambda}+\delta\hat{\lambda
}(\hat{A},\hat{\lambda},\delta\theta)+O(\delta\theta^{2})$ and expand
(\ref{compact}) to first order in $\delta\theta$. It then follows from
(\ref{var}) that
\begin{equation}
\delta_{\tilde{\lambda}}\delta\hat{A}_{i}-\partial_{i}\delta\hat{\lambda
}-i[\delta\hat{\lambda},\hat{A}_{i}]_{\ast}-i[\delta\hat{A}_{i},\hat{\lambda
}]_{\ast}=-\frac{1}{2}\delta\theta^{kl}\{\partial_{k}\hat{A}_{i},\partial
_{l}\hat{\lambda}\}_{\ast}. \label{1stord}%
\end{equation}
Following the discussion of \cite{6}, one gets a particular solution%
\begin{align}
&  \delta\hat{A}_{i}=-\frac{1}{4}\delta\theta^{kl}\{\hat{A}_{k},\partial
_{l}\hat{A}_{i}+\hat{F}_{li}\}_{\ast},\nonumber\\
&  \delta\hat{\lambda}=\frac{1}{4}\delta\theta^{kl}\{\partial_{k}\hat{\lambda
},\hat{A}_{l}\}_{\ast}. \label{par}%
\end{align}
Notice, however, that in the above process one gets two unknowns from a single
equation (\ref{1stord}), so naturally there might be other solutions which
solve (\ref{1stord}) just as well. In ref.\cite{13}, such redundant solutions
were indeed found (and referred to as ambiguities of the Seiberg-Witten map),
which takes the form%
\begin{align}
&  \delta\hat{A}_{i}=-\frac{1}{4}\delta\theta^{kl}\{\hat{A}_{k},\partial
_{l}\hat{A}_{i}+\hat{F}_{li}\}_{\ast}+\alpha\delta\theta^{kl}\hat{D}_{i}%
\hat{F}_{kl}+\beta\delta\theta^{kl}\hat{D}_{i}[\hat{A}_{k},\hat{A}_{l}]_{\ast
},\nonumber\\
&  \delta\hat{\lambda}=\frac{1}{4}\delta\theta^{kl}\{\partial_{k}\hat{\lambda
},\hat{A}_{l}\}_{\ast}+2\beta\delta\theta^{kl}[\partial_{k}\hat{\lambda
},\hat{A}_{l}]_{\ast}, \label{ambia}%
\end{align}
where $\alpha,\beta$ are some arbitrary commuting constant parameters. The
$\alpha,\beta$ dependent terms in $\delta\hat{A}_{i}$ can be regarded as a
field dependent gauge transformation, and hence can be neglected because
Seiberg-Witten map is an identification between gauge equivalent classes,
rather than identification between field configurations.

However, it was argued also in \cite{13} that there are further ambiguities
arising from successive applications of Seiberg-Witten maps, for which part
can be absorbed by a field-dependent gauge transformation, and the rests could
be removed by a field redefinition. We shall make much detailed discussion on
this point for the case of gauge theory coupled with matter and find some
different conclusion.

\section{Ambiguities of Seiberg-Witten map in the presence of matter field}

After making a brief review and setting up our notations, now we come to our
central subject -- Seiberg-Witten map in the presence of fermionic matter
field. Assume now that, in addition to the purely gauge field configuration
described in (\ref{gaugefield}), our theory contains also some fermionic
matter field $\psi$ which transforms under the fundamental representation of
the gauge group,%
\begin{equation}
\delta_{\hat{\lambda}}\psi=i\hat{\lambda}\ast\psi. \label{matter}%
\end{equation}
Assume also that the basic concept of Seiberg-Witten map still holds in the
presence of matter field, which means that the gauge equivalent classes for
two such theories with small deviation $\delta\theta$ in the noncommutativity
parameter can still be identified. Then, in addition to the standard
Seiberg-Witten map (\ref{SW},\ref{compact}) for gauge fields, we have also the
following identities for the matter field,%
\begin{align}
\tilde{\psi}(\hat{\psi})+\delta_{\tilde{\lambda}}\tilde{\psi}(\hat{\psi
},\hat{A})  &  =\tilde{\psi}(\hat{\psi}+\delta_{\tilde{\lambda}}\hat{\psi
},\hat{A}+\delta_{\hat{\lambda}}\hat{A}),\label{SW2}\\
\delta_{\tilde{\lambda}}\delta\hat{\psi}  &  =\delta\delta_{\tilde{\lambda}%
}\hat{\psi}. \label{compact2}%
\end{align}

In order to solve the above equations, we write, just as in the case of purely
gauge theory, the following expansion for the matter field,%
\begin{equation}
\tilde{\psi}=\hat{\psi}+\delta\hat{\psi}(\hat{\psi},\hat{A},\hat{\lambda
},\delta\theta)+O(\delta\theta^{2}).\label{psi}%
\end{equation}
Substituting (\ref{psi}) into (\ref{compact2}), we get%
\begin{equation}
\delta_{\hat{\lambda}}\delta\hat{\psi}-i\hat{\lambda}\ast\delta\hat{\psi
}-i\delta\hat{\lambda}\ast\hat{\psi}=-\frac{1}{2}\delta\theta^{kl}\partial
_{k}\hat{\lambda}\ast\partial_{l}\hat{\psi}.\label{1stord2}%
\end{equation}
It can be easily checked that(See Refs.\cite{3})
\begin{equation}
\delta\hat{\psi}=-\frac{1}{2}\delta\theta^{kl}\hat{A}_{k}\ast\partial_{l}%
\hat{\psi}+\frac{i}{4}\delta\theta^{kl}\hat{A}_{k}\ast\hat{A}_{l}\ast\hat
{\psi}\label{psi2}%
\end{equation}
is a particular solution to (\ref{1stord2}).

Now let us consider the ambiguities arisen in the presence of matter fields.
According to equation (\ref{1stord2}), suppose there exits some functions
$\hat{\psi}^{\prime},\hat{\lambda}$ such that
\begin{equation}
\delta_{\hat{\lambda}}\hat{\psi}^{\prime}-i\hat{\lambda}\ast\hat{\psi}%
^{\prime}-i\hat{\lambda}^{\prime}\ast\hat{\psi}=0, \label{ambi}%
\end{equation}
then $\delta\hat{\psi}+\hat{\psi}^{\prime},\delta\hat{\lambda}+\hat{\lambda
}^{\prime}$ also solve (\ref{1stord2}). Therefore, any solution $\hat{\psi
}^{\prime},\hat{\lambda}^{\prime}$ of (\ref{ambi}) would give rise to
ambiguities for the Seiberg-Witten map (\ref{SW},\ref{SW2}). After some
algebra, we find that the solution to (\ref{ambi}) is%
\begin{align*}
\hat{\lambda}^{\prime}  &  =2\beta\delta\theta^{kl}[\partial_{k}\hat{\lambda
},\hat{A}_{l}]_{\ast},\\
\hat{\psi}^{\prime}  &  =i\alpha\delta\theta^{kl}\hat{F}_{kl}\ast\hat{\psi
}+i\beta\delta\theta^{kl}[\hat{A}_{k},\hat{A}_{l}]_{\ast}\ast\hat{\psi}.
\end{align*}
Combining with the corresponding result for the case of gauge fields (see eq.
(\ref{ambia}) of the last section), we see that the $\alpha,\beta$ ambiguities
in $\delta\hat{A}$ and $\delta\hat{\psi}$ are both of the form of a gauge
transformation with gauge parameter $\alpha\delta\theta^{kl}\hat{F}_{kl}%
+\beta\delta\theta^{kl}[\hat{A}_{k},\hat{A}_{l}]_{\ast}$ and therefore can be
neglected just as in the case of a pure gauge theory.

Next we come to the case when more ambiguities may arise from successive
applications of Seiberg-Witten maps. Consider for instance that we wish to map
a theory with noncommutativity parameter $\theta$ to a theory with
noncommutativity parameter $\theta+\delta\theta_{1}+\delta\theta_{2}$. Of
cause we can realize the map by a two-step process: we may either map from
$\theta$ to $\theta+\delta\theta_{1}$ and then to $\theta+\delta\theta
_{1}+\delta\theta_{2}$ or from $\theta$ to $\theta+\delta\theta_{2}$ and then
to $\theta+\delta\theta_{1}+\delta\theta_{2}$. Naively both paths should work
the same way. However, a detailed check reveals that the two paths are
actually non-equivalent.

Let us now study the details of the new ambiguities arisen from the
non-equivalence of the two paths just stated. Along the two paths the fields
are mapped as%
\[%
\begin{array}
[c]{l}%
\theta\rightarrow\theta+\delta\theta_{1}\rightarrow\theta+\delta\theta
_{1}+\delta\theta_{2}\\
\hat{\psi}\rightarrow\tilde{\psi}\rightarrow\overset{\eqsim}{\psi},\\
\hat{A}\rightarrow\tilde{A}\rightarrow\overset{\eqsim}{A};
\end{array}
\]%
\[%
\begin{array}
[c]{l}%
\theta\rightarrow\theta+\delta\theta_{2}\rightarrow\theta+\delta\theta
_{1}+\delta\theta_{2}\\
\hat{\psi}\rightarrow\bar{\psi}\rightarrow\overset{\simeq}{\psi},\\
\hat{A}\rightarrow\bar{A}\rightarrow\overset{\simeq}{A}.
\end{array}
\]

For the first path, equation (\bigskip\ref{psi2}) yields%
\[
\overset{\eqsim}{\psi}=\tilde{\psi}-\frac{1}{2}\delta\theta_{2}^{kl}%
\tilde{A}_{k}\ast^{\prime}\partial_{l}\tilde{\psi}+\frac{i}{4}\delta\theta
_{2}^{kl}\tilde{A}_{k}\ast^{\prime}\tilde{A}_{l}\ast^{\prime}\tilde{\psi},
\]
where $\ast^{\prime}$ is the Moyal product defined with the noncommutativity
parameter $\theta+\delta\theta_{1}$. Applying (\bigskip\ref{psi2}) again, we
get%
\begin{align}
\overset{\eqsim}{\psi}  &  =\tilde{\psi}-\frac{1}{2}(\delta\theta_{1}%
+\delta\theta_{2})^{kl}\tilde{A}_{k}\ast\partial_{l}\tilde{\psi}+\frac{i}%
{4}(\delta\theta_{1}+\delta\theta_{2})^{kl}\tilde{A}_{k}\ast\tilde{A}_{l}%
\ast\tilde{\psi}\nonumber\\
&  +\frac{1}{4}\delta\theta_{2}^{kl}\delta\theta_{1}^{pq}\left[
\hat{A}_{k}\ast\partial_{l}(\hat{A}_{p}\ast\partial_{q}\hat{\psi})-\frac{i}%
{2}\hat{A}_{k}\ast\partial_{l}(\hat{A}_{p}\ast\hat{A}_{q}\ast\hat{\psi
})\right. \nonumber\\
&  +\frac{1}{2}\{\hat{A}_{p},\partial_{q}\hat{A}_{k}+\hat{F}_{qk}\}_{\ast}%
\ast\partial_{l}\hat{\psi}-i\partial_{p}\hat{A}_{k}\ast\partial_{q}%
\partial_{l}\hat{\psi}\nonumber\\
&  -\frac{i}{4}\hat{A}_{k}\ast\{\hat{A}_{p},\partial_{q}\hat{A}_{l}+\hat
{F}_{ql}\}_{\ast}\ast\hat{\psi}-\frac{i}{4}\{\hat{A}_{p},\partial
_{q}\hat{A}_{k}+\hat{F}_{qk}\}_{\ast}\ast\hat{A}_{l}\ast\hat{\psi}\nonumber\\
&  -\frac{i}{2}\hat{A}_{k}\ast\hat{A}_{l}\ast\hat{A}_{p}\ast\partial_{q}%
\hat{\psi}-\frac{1}{4}\hat{A}_{k}\ast\hat{A}_{l}\ast\hat{A}_{p}\ast
\hat{A}_{q}\ast\hat{\psi}\nonumber\\
&  -\frac{1}{2}\partial_{p}\hat{A}_{k}\ast\partial_{q}\hat{A}_{l}\ast\hat
{\psi}-\frac{1}{2}\partial_{p}\hat{A}_{k}\ast\hat{A}_{l}\ast\partial_{q}%
\hat{\psi}\nonumber\\
&  -\left.  \frac{1}{2}\hat{A}_{k}\ast\partial_{p}\hat{A}_{l}\ast\partial
_{q}\hat{\psi}\right]  . \label{delta2}%
\end{align}
The same procedure for the second path would give rise corresponding result
for $\overset{\simeq}{\psi}$, which is basically of the same form as
$\overset{\eqsim}{\psi}$, but with $\delta\theta_{1}$ and $\delta\theta_{2}$
exchanged:%
\[
\overset{\simeq}{\psi}=\overset{\eqsim}{\psi}(\mathrm{with}\quad\delta
\theta_{1}\leftrightarrow\delta\theta_{2}).
\]
Therefore, to first order in $\delta\theta$, $\overset{\simeq}{\psi}$ and
$\overset{\eqsim}{\psi}$ are equal, while to the order of $\delta\theta
_{1}\delta\theta_{2}$, we get%
\begin{align}
\lbrack\delta_{1},\delta_{2}]\hat{\psi}  &  \equiv\overset{\eqsim}{\psi
}-\overset{\simeq}{\psi}\nonumber\\
&  =-\frac{i}{4}\delta\theta_{2}^{kl}\delta\theta_{1}^{pq}\hat{F}_{pk}\ast
\hat{D}_{l}\hat{D}_{q}\hat{\psi}\nonumber\\
&  -\frac{1}{8}\delta\theta_{2}^{kl}\delta\theta_{1}^{pq}\left[  -2\hat
{F}_{pk}\ast(\partial_{l}\hat{A}_{q}+i\hat{A}_{l}\ast\hat{A}_{q})\right.
\nonumber\\
&  +i\hat{A}_{k}\ast\partial_{l}(\hat{A}_{p}\ast\hat{A}_{q})-i\hat{A}_{p}%
\ast\partial_{q}(\hat{A}_{k}\ast\hat{A}_{l})\nonumber\\
&  +\partial_{p}\hat{A}_{k}\ast\partial_{q}\hat{A}_{l}-\partial_{k}%
\hat{A}_{p}\ast\partial_{l}\hat{A}_{q}\nonumber\\
&  +\frac{i}{2}[\hat{A}_{k},\{\hat{A}_{p},\partial_{q}\hat{A}_{l}+\hat{F}%
_{ql}\}_{\ast}]_{\ast}\nonumber\\
&  -\frac{i}{2}[\hat{A}_{p},\{\hat{A}_{k},\partial_{l}\hat{A}_{q}+\hat{F}%
_{lq}\}_{\ast}]_{\ast}\nonumber\\
&  +\left.  \frac{1}{2}[\hat{A}_{k}\ast\hat{A}_{l},\hat{A}_{p}\ast
\hat{A}_{q}]_{\ast}\right]  \ast\hat{\psi}. \label{ambi2}%
\end{align}
In the last calculations, we did not consider the $\alpha,\beta$ dependent
ambiguities which were thought to be trivial. In fact, a careful check on the
$\alpha,\beta$ dependent terms would show that their contribution to the
expression $[\delta_{1},\delta_{2}]\hat{\psi}$ is still of the form
($\alpha,\beta$ dependent terms)$\ast\hat{\psi}$. So the $\alpha,\beta$
dependent ambiguities are indeed only a matter of gauge choice and hence can
be neglected. The rest parts of $[\delta_{1},\delta_{2}]\hat{\psi}$ are
grouped into two terms -- the first term is a gauge covariant ``local''
differential polynomial in the fundamental fields $\hat{A}$ and $\hat{\psi}$,
the second term again takes the form of a field-dependent gauge transformation
over $\hat{\psi}$.

At first sight, one may tends to make the same conclusion as in \cite{13} that
one part of the last ambiguities may be absorbed by a gauge transformation and
the other part by a local field redefinition. But actually this is not the
case. Let us recall that the action of $[\delta_{1},\delta_{2}]$ on $\hat{A}$
was already given in ref. \cite{13},%
\begin{align}
\lbrack\delta_{1},\delta_{2}]\hat{A}_{i}  &  =\frac{i}{8}\delta\theta_{2}%
^{kl}\delta\theta_{1}^{pq}[\hat{F}_{kp},\hat{D}_{l}\hat{F}_{qi}+\hat{D}%
_{q}\hat{F}_{li}]_{\ast}\nonumber\\
&  +\frac{1}{16}\delta\theta_{2}^{kl}\delta\theta_{1}^{pq}\hat{D}_{i}\left[
\frac{1}{2}\{\hat{A}_{k},\{\hat{A}_{p},\hat{F}_{lq}\}_{\ast}\}_{\ast}+\frac
{1}{2}\{\hat{A}_{p},\{\hat{A}_{k},\hat{F}_{lq}\}_{\ast}\}_{\ast}\right.
\nonumber\\
&  +\left.  \frac{1}{2}[[\hat{A}_{k},\hat{A}_{p}]_{\ast},\partial
_{l}\hat{A}_{q}+\partial_{q}\hat{A}_{l}]_{\ast}-i[\partial_{p}\hat{A}_{k}%
,\partial_{l}\hat{A}_{q}]_{\ast}+i[\partial_{k}\hat{A}_{p},\partial
_{q}\hat{A}_{l}]_{\ast}\right] \nonumber\\
&  +\hat{D}_{i}(\alpha,\beta\quad\mathrm{dependent\quad terms}). \label{ambi4}%
\end{align}

Apart from the terms which are ``local'', gauge covariant differential
polynomials of the fundamental fields, the terms which take the form of field
dependent gauge transformations in eqs. (\ref{ambi2}) and (\ref{ambi4}) have
different gauge parameters. Consequently, although one might be able to cancel
all the terms on the right hand side of (\ref{ambi4}) by simultaneous use of
gauge transform and field redefinition, one cannot simultaneously remove all
the terms appeared in (\ref{ambi2}) by the same gauge transform. In other
words, if initially $\hat{A}_{i}$ and $\hat{\psi}$ are in the same gauge slice
of the theory with noncommutativity parameter $\theta$, the successive
application of two Seiberg-Witten maps along two different paths in ``$\theta
$-space'' would not lead to field configurations which live in the same gauge
slice in the resulting theory. This path dependence in the ``$\theta$-space''
might imply that there exist some nontrivial topological obstacles in the
space of noncommutativity parameter which prevents one from deforming one
noncommutative gauge theory coupled with matter into another in a smooth way.
Perhaps a relatively safe manner in studying noncommutative field theories is
to keep the noncommutativity parameter fixed, before we have a complete
understanding of the mysterious ambiguities described above.

\end{document}